\documentstyle[12pt,epsf]{article}

\begin{document} 
 
\title{The family of regular interiors for non-rotating black holes with 
$T^0_0 = T^1_1$} 
\author{Emilio Elizalde$^{\cite{email,http}}$ and 
Sergi R. Hildebrandt$^{ \cite{email2}}$,\footnote{Temporary address: Avda. 
Mar\'\i tima, 39. P-041E, Candelaria, S/C. de Tenerife, Spain.} 
\\Instituto de Ciencias del Espacio (CSIC) \ \& \\ 
Institut d'Estudis Espacials de Catalunya (IEEC/CSIC) \\ 
Edifici Nexus, Gran Capit\`a 2-4, 08034 Barcelona, Spain} 
\date{} 
\maketitle 
 
 
\newcommand{\rf}[1]{{\rm (\ref{#1})}} 
 
\def\ab{\alpha \beta} 
\def\lh{\hbox{{\boldmath \hbox{$ \ell $}}} } 
\def\th{\hbox{{\boldmath \hbox{$ \Theta $}}} }      
\def\l{\Lambda} 
\def\le{\Lambda_1} 
\def\li{\Lambda_2} 
\def\tr{{\tilde r} } 
\def\dal{\sqcap_{\smash{\hskip -5.7pt \lower 1.3pt \hbox{--}}}\,\! }  
\def\ks{Kerr-Schild } 
 
\def\bce{\begin{center}} 
\def\ece{\end{center}} 
\def\beq{\begin{eqnarray}} 
\def\eeq{\end{eqnarray}} 
\def\ben{\begin{enumerate}} 
\def\een{\end{enumerate}} 
\def\ul{\underline} 
\def\ni{\noindent} 
\def\nn{\nonumber} 
\def\bs{\bigskip} 
\def\ms{\medskip} 
 
\begin{abstract} 
 We find the general solution for the spacetimes describing the 
interior of static black holes with an equation of state of the type $ T^0_0 
= T^1_1 $ ($T$ being the stress-energy tensor). This form is the one expected from taking into account different quantum effects associated with strong gravitational fields. 
We recover all the particular examples found in the literature. We 
remark that all the 
solutions  found follow the natural scheme of an interior core linked smoothly 
with the exterior solution by a transient region. We also discuss 
 their local 
energy properties and give the main ideas involved in a possible 
generalization of the scheme, in order to include other 
realistic types of sources. 
\end{abstract} 
\section{Introduction} 
\label{s-int} 
Any static black hole (BH) arises from the gravitational collapse of some 
object. Under the premises in this work, the object has not (yet) 
shrinked indefinitely 
and has not given rise to a spacetime singularity. It is then natural 
to consider 
two regions: one {\em exterior} to the object, and the object {\em itself}. The 
exterior region, as is well known, can be described by a spacetime belonging 
to the Reissner-Nordstr\"om \cite{mtw,chandra} solution. In the absence of 
electric ---or magnetic--- charge it is simply Schwarzschild's spacetime. 
Furthermore, one can add a cosmological constant, following recent 
observational results \cite{perl,ries}. Then, the spacetime belongs to the 
Kottler-Trefftz solution family \cite{kottler,trefftz}. 
In this case, the global 
properties of the spacetime clearly change, e.g., the spacetime is no longer 
asymptotically flat (see e.g. \cite{mtw,he}). 
 
Going a step further, we consider the body itself as composed of 
two main regions. One is 
its {\em surface} and the other the rest of the body, i.e. the {\em 
interior} region. One may expect 
that some mechanism ---having to do e.g. with quantum 
gravity--- will be able to stop the collapse of the body. Therefore, 
we will think of the interior of the body as being described by 
some spacetime product of the present knowledge 
available on the merge of quantum field 
theory and gravitation. A widely studied issue in this direction 
 is that of quantum vacuum 
effects \cite{casimir}--\cite{eli1} and the resulting spacetime 
turns out to be  a de Sitter (dS) 
or anti de Sitter (AdS) one \cite{glinner}--\cite{bardeen}. There 
are other alternatives, as e.g., those of \cite{ayon2,ayon3}. 
 
In all these cases considered so far, either no distinction has 
been made between the interior and 
the exterior of the body (see e.g., \cite{dymni}--\cite{ayon3}) 
 or there 
appears a {\em singular} distribution of matter at the surface of 
the body (see e.g. \cite{ip,fmm}). This distribution is singular 
in the sense that 
it is a {\em matter surface density} ---called {\em singular shell}. 
However, contrary to the case of electromagnetic charge densities, a matter 
surface density has neither been observed, nor is it predicted by any theory. 
It is 
thus more natural to assume that the matter on the surface of the body is 
distributed across the body, and leave for a subsequent 
 study the issue of whether this region 
is thin or thick, in comparison with the region dominated by quantum vacuum 
effects, through one of the solutions referred to before. Finally, the only 
work considering {\em all} the features of the structure of a regular static 
BH with a clear physical source is \cite{magli1}. However, Nariai spacetime was absent, 
as well as an implementation of previous attempts and a complete study of (local) energy conditions. 
Thus, in our opinion, it is worth carrying out a unification of the 
different 
results obtained so far, as well as extending them in order to cover some impotant issues 
 that were overlooked  in those analysis. 
Here we provide, for the first time to our knowledge, the 
general solution of 
the scheme discussed above. In particular, we carry out 
 an implementation of all those previous works which, for one or 
another reason did not comply with all the requirements already 
specified. 
We also perform a study of the local energy conditions in all 
these cases. 
 
Finally, it is also important to introduce other kind of solutions for the 
interior region, aside from the ones referred to before, which arise 
from results, or 
just hints, coming from the contribution of the quantum vacuum to gravity. 
To summarize, these are the points that will be dealt with, 
successively,  in the body 
of this paper comprising the next 10 sections. They are clearly 
identified by their titles and will need no further specification here. 
Sect.~\ref{s-fr} is devoted to some final remarks, 
 and in Sect.~\ref{s-c} we 
provide the conclusions of the work. A brief survey 
can be found in \cite{rqibh}. 
 
Throughout this work we will use units such that $ G = 1 $, $ c = 1 $, 
Einstein's equations are written in the form $ G_{\ab} = 8 \pi T_{\ab} $, 
where $ G_{\ab} $ is the Einstein tensor ---we follow the conventions of 
\cite{mtw}--- and $ T_{\ab} $ is the energy-momentum tensor. A prime will 
denote derivation with respect to the coordinate $ r $. 
\section{Spacetimes with a SSQV as a source} 
\label{s-swa} 
Due to the imposed limitation of non-accepting singular mass shells, 
the spacetimes 
describing the interior of the body are not allowed to be of some well 
known kind as 
dS or AdS spacetimes. Indeed, the interior solution cannot be everywhere a 
spatial isotropic solution, as dS or AdS, because this would lead to a 
sudden change in the pressures exerted by the body to the exterior, and 
would lead to the appearance of a singular mass shell~\cite{israel2}. Now, 
the type of generalization depends on the underlying physics one is able to 
assume. The isotropic case is suitable in order to link it with the expected 
contributions of a dominating quantum vacuum, especially those associated 
with vacuum polarization. 
As we are dealing with spacetimes which are spherically symmetric, a natural 
generalization is to assume that the body may be described by a solution 
which is {\em invariant} to any non-rotating observer, with a free radial 
motion, instead of a solution which is invariant to any observer. This 
generalization of the energy-matter content of the body is called {\em 
spherically symmetric quantum vacuum} (SSQV), after \cite{dymni} ---see also 
\cite{dymni2}--- and requires the imposition of $ T^0_0 = T^1_1 $, for any 
non-rotating observer. This is the type of enery-matter content that is 
considered in \cite{ip}--\cite{bronnikov} and will be the one used in 
the first part of this work,  until we get to  Sect.~\ref{s-tmo}. 
In particular, SSQVs and non-linear electrodynamics 
have given some relevant results on the issue of regular BHs, see 
e.g.~\cite{ayon}--\cite{ayon2}, \cite{bronnikov}. 
 
We shall now characterize the families of spacetimes that are suitable to 
become SSQVs. 
Any static, spherically symmetric spacetime can be conveniently 
described by 
\beq 
\label{met1} 
ds^2 = -F(r) \, dt^2+ F^{-1}(r) \, dr^2 + G^2(r) \, d\Omega^2, 
\eeq 
where $ d\Omega^2 \equiv d\theta^2 + \sin^2\theta \, d\varphi^2 $. 
There 
are certainly other ways to represent these spacetimes ---which avoid the 
problems occurring near the 
possible horizons--- or by putting $ R^2 d\Omega^2 $, provided 
$ G' \equiv  dG(r)/dr \neq 0 $ (see e.g. \cite{msi,wheeler}). 
For a local 
observer at rest with respect to the coordinate grid of~\rf{met1}, a 
standard computation of $ T_{\ab} $ yields ($ \rho $ is the energy density, 
$ p $ the radial pressure and $ p_2 $, $ p_3 $ the tangential pressures, 
measured by this observer) 
\beq 
\label{eq-r} 
8 \pi \rho = {1 \over G^2} \Bigl[ 1 - F(G'^2 + 2 G G'') - G G' F' \Bigr], \\ 
\label{eq-p} 
8 \pi p = {1 \over G^2} \Bigl[ -1 + F G'^2 + G G' F' \Bigr], \\ 
8 \pi p_2 = 8 \pi p_3 = {F'' \over 2} + {F G'' \over G} + {F' G' \over G}. 
\eeq 
Imposing $ \rho + p = 0 $ in Eqs.~\rf{eq-r},~\rf{eq-p}, we get 
\beq 
F \, G'' = 0. 
\eeq 
We now use that $ G $ cannot be zero in any open region. Two 
alternatives appear: $ F = 0 $ or $ G'' = 0 $. If $ F = 0 $, the 
expression~\rf{met1} is useless. It is first necessary to change 
the coordinate system of~\rf{met1} by $ dT \equiv dt + (1-F)/F dr $, 
while keeping the rest unchanged. Then one can impose $ F = 0 $. The result 
is 
\beq 
\label{eq-fn0} 
ds^2 = 2 \, dT \, dr + 2 \,dr^2 + G^2(r) \, d\Omega^2. 
\eeq 
In the orthonormalized cobasis given by $ \th^0 = dT/\sqrt{2} $, $ \th^1 = 
dT/\sqrt{2} + \sqrt{2} \, dr $, $ \th^2 = G \, d\theta $, $ \th^3 = G 
\sin\theta \, d\varphi $, the Ricci tensor takes the form 
\beq 
\begin{array}{ll} 
\label{ricci0} 
\hbox{\bf Ricci} = & {\displaystyle {G'' \over G}} ( - \th^0 \otimes \th^0 + 
\th^0 \otimes \th^1 + \th^1 \otimes \th^0 - \th^1 \otimes \th^1) \cr 
& + \ {\displaystyle{1 \over G^2}} ( \th^2 \otimes \th^2 + \th^3 \otimes 
\th^3). 
\end{array} 
\eeq 
On the other hand, for a SSQV we must have $ \rho + p = 0 $ (the conditions 
on $ T_{\ab} $ being directly translated into conditions for $ R_{\ab} $) 
\beq 
\label{ricci} 
\hbox{\bf Ricci} = R_{00}( - \th^0_N \otimes \th^0_N + \th^1_N \otimes 
\th^1_N) 
+ R_{22} ( \th^2_N \otimes \th^2_N + \th^3_N \otimes \th^3_N), 
\eeq 
where $ \{ \th^{\Omega}_N \} $ is some orthonormalized cobasis, 
not necessarily coincident with the one used in the computation 
of~\rf{ricci0}. Therefore, we must look for an orthonormalized cobasis for 
which the Ricci tensor~\rf{ricci0} becomes of the type~\rf{ricci}. Clearly 
this is the same as finding out whether we can have linear expressions 
$ \th^0_N \equiv A \th^0 + B \th^1 $, and 
$ \th^1_N = C \th^0 + D \th^1 $, with 
$ -\th^0_N \cdot \th^0_N =  \th^1_N \cdot \th^1_N = 
\th^0_N \cdot \th^1_N + 1= 1 $. However, Eq.~\rf{ricci} is invariant under 
these changes. The only solution that makes~\rf{ricci0} and~\rf{ricci} 
compatible is then 
\beq 
G'' = 0. 
\eeq 
If $ F \neq 0 $, we also have $ G'' = 0 $. Thus $ G'' = 0 $ constitutes 
the {\it proper} characterization of any possibility. 
 
Now, from $ G'' = 0 $ two distinct alternatives appear 
\beq 
G = \gamma , \quad \hbox{or} \quad G= \alpha \, r + \gamma, 
\eeq 
where $ \alpha (\neq 0) $ and $ \gamma $ are constant. Only the latter has 
been considered in detail in the literature of regular BHs. We will study it 
in the sequel. 
\subsection{Other expressions for the spacetimes describing SSQVs} 
In order to include the possible horizons, we write the 
metrics~\rf{met1} under the common form 
\beq 
\label{met2} 
ds^2 = - (1-H) \, dT^2 + 2 H \, dT \, dr + ( 1 + H ) \, dr^2 
+ \gamma^2 \, d\Omega^2, 
\\ 
\label{met3} 
ds^2 = - (1-H) \, dT^2 + 2 H \, dT \, dr + ( 1 + H ) \, dr^2 
+ r^2 \, d\Omega^2, 
\eeq 
where $ H \equiv 1 - F $, and the coordinate change is given by $ dT = dt + 
(1-F)/F dr $. We will use these forms in the sequel. We have also used the 
fact that the case $ g= \alpha r + \gamma $ is physically equivalent to the 
case $ G = r $. This is intuitively seen because $ \alpha $ merely 
represents the scale of units used for $ r $ and $ \gamma$ is an arbitrary 
(constant) origin. In terms of coordinate changes we have: The 
metric~\rf{met1} for $ G = \alpha r + \gamma $ is $ ds^2 = -F(r) \, dt^2+ 
F^{-1}(r) \, dr^2 + (\alpha r + \gamma )^2 \, d\Omega^2 $. Recalling that $ 
\alpha \neq 0 $, one can define a new radial coordinate $ \tr \equiv \alpha 
r + \gamma $. The metric becomes then $ ds^2 = -F[(\tr - \gamma)/\alpha] \, 
dt^2+ \alpha^{-2}F^{-1}[(\tr - \gamma)/\alpha] \, d\tr^2 + \tr^2 \, 
d\Omega^2 $. Now, under a reparametrization of the $ t $ coordinate by $ dt 
\equiv \alpha d {\tilde t} $ we get 
$ ds^2 = -\alpha^2 F[(\tr - \gamma)/\alpha] \, d{\tilde t}^2+ 
\alpha^{-2}F^{-1}[(\tr - \gamma)/\alpha] \, d\tr^2 + \tr^2 \, d\Omega^2 $. 
Whence one can conclude that any member of~\rf{met1} with $ G = \alpha r + 
\gamma $ with $ \alpha \neq 0 $ is equivalent to another member of~\rf{met1} 
with $ \alpha = 1 $ and $ \gamma = 0 $. Since we are studying the general 
description of SSQVs it is enough to consider the representation $ \alpha = 
1$, $ \gamma = 0 $ and {\em arbitrary} $ F (r) $ to include {\em any} case 
of SSQV. 
 
Furthermore, it is easy to show that~\rf{met3} can be written in the 
Kerr-Schild form \cite{ks}: 
\beq 
ds^2 = ds^2_{\eta} + 2 H(r) \lh \otimes \lh, 
\eeq 
where $ ds^2_{\eta} $ stands for the 
flat spacetime metric, $ H $ is an arbitrary function of $ r $ 
and $ \lh $ is a geodesic radial null one-form, $ \lh = (1/\sqrt{2})(dt 
\pm\, dr) $. Thus, the SSQVs in~\rf{met3} can be thought as the 
family of maximal spherically symmetric spacetimes expanded by a 
geodesic radial null one-form from flat spacetime (GRNSS spaces). 
 
To summarize, there are {\it only} two ---non-equivalent--- families 
of SSQV. The case with $ G' = 0 $ is characteristic of the Nariai 
solution \cite{nariai,kramer}. The Nariai solution is a solution of 
Einstein's equations for the same pattern as the de Sitter 
solution, i.e. $ T_{\ab} = \Lambda_0 g_{\ab} $, being $ \Lambda_0 $ 
the cosmological constant. The difference lies in the ``radial'' coordinate. 
In the Nariai case there is no proper center for the spherical symmetry. 
Therefore, we shall call the spacetimes with $ G' =0 $ generalized 
Nariai metrics. Finally, the other case corresponds to the GNRSS which 
constitute a distinguished family of the class of Kerr-Schild metrics. 
\section{Geometrical properties of the solutions} 
\label{s-gps} 
\subsection{Generalized Nariai metrics} 
\label{ss-gnm} 
Using an orthonormal cobasis defined as $ \th^0 = (1-{H \over 2} ) \, dT - 
{H \over 2} \, dr $, $ \th^1 = ( 1 + {H \over 2} )\, dr + {H \over 2} \, dT 
$, $ \th^2 = \gamma \, d \theta $, $ \th^3 = \gamma \, \sin\theta \, 
d\varphi $, we see that the Riemann tensor has as independent components 
\beq 
R_{0101} = -{H'' / 2}, \quad R_{2323} = {1 / \gamma^2}. 
\eeq 
The Ricci tensor is characterized by $ R_{00} = - R_{11} = -{H'' / 2} $, $ 
R_{22} =  R_{33} = {1 / \gamma^2} $. 
The scalar curvature is $ R = H'' + 2/\gamma^2 $, and the Einstein 
tensor has the following non-zero components 
\beq 
G_{00}= - G_{11} = {1 / \gamma^2} , \quad G_{22} = G_{33} = - {H'' / 2}, 
\eeq 
The isotropic solution ---the one to be found at the core--- 
yields $ H = (1/\gamma^2) r^2 + b r + c $, where $ b $ and $ c $ are 
arbitrary constants. Without losing generality, 
we can set $ b $, $ c  = 0 $ (as they are clearly gauge freedoms for any 
spacetime in the family). Thanks to the presence of the Nariai solution 
inside this family ---$ T^{\rm Nariai}_{\ab} = \Lambda_0 g_{\ab} $ --- the 
factor $ 1/\gamma^2 $ can be identified with $ \Lambda_0 $. Thus, the {\it 
only isotropic} quantum vacuum belonging to this family is the Nariai 
solution. 
\subsection{The GNRSS metrics} 
\label{ss-tgn} 
First we note that these spacetimes fulfill the relation 
\beq 
\label{pr} 
p_2 = -\rho - {\rho'\, r \over 2}. 
\eeq 
As a consequence, for a regular source $ p_2 \to - \rho $ as $ r = 0 $ is 
approached. Therefore, in {\em any regular solution} the spacetime becomes 
more and more isotropic as $ r \to 0 $. Thus, the contribution from the 
quantum  vacuum becomes more and more 
dominant as $ r $ tends to 0 and the initial idea of distributing the 
singular mass shell across some region of the body is completed. 
 
On the other hand, one can choose a similar cobasis as in the Nariai-like 
case, just replacing  $ \gamma $ by $ r $. The Riemann tensor has the 
following independent components, 
\beq 
R_{0101} = - {{H''}\over 2} , \quad R_{0202} = R_{0303} = -{{H'} \over 2r} 
,  \cr R_{1212} = R_{1313} = {{H'} \over 2r} , \quad 
R_{2323} = { H \over  r\sp{2}}. 
\eeq 
The Ricci tensor for these spacetimes has the following non-zero components 
$ R_{00} = -R_{11} = - (1/2) [ H'' + (2 H'/ r ) ]$, $ R_{22} = R_{33} = (1 / 
r) [ H' + (H / r) ] $. 
And the scalar curvature is given by $ R = H'' + (4 H'/ r) + (2 H / r\sp{2}) 
$. 
Finally, the non-zero components of the Einstein's are $ 
G_{00} = -G_{11} = ( 1 / r ) [ H' + ( H / r)] $, $ G_{22} = G_{33} = - (1 / 
2) [ H'' + (2 H' / r) ] $. 
Other expressions that will be used later are 
\beq 
\label{eqe} 
G_{00} = - G_{11} = {1 \over r^2} (H r)', \quad G_{22} = G_{33} = G_{00} - 
{G_{00}' \ r \over 2} . 
\eeq 
 
In this case, the isotropic (regular) GNRSS is the de Sitter solution, given 
by $ H(r) = (\Lambda_0/3) r^2 $. 
\subsubsection{Exterior metrics and GNRSS metrics} 
\label{s-emg} 
It turns out that {\em all} of the possible exterior metrics ---see 
Sect.~\ref{s-int}--- also belong to the GNRSS family. The function $ H $ is 
$ H (r) = (\Lambda_{\rm ext}/3) r^2 + 2m/r - q^2/r^2 $, where $ \Lambda_{\rm 
ext} $ stands for the external cosmological constant, $ m $ is the ADM mass 
of the BH and $ q $ its electromagnetic charge. 
 
This coincidence will be very useful in the following section. 
 
\section{Junction of the interior and exterior solutions} 
\label{s-joe} 
The junction, or matching, of two spherically symmetric spacetimes is 
well-known (see e.g. ~\cite{israel2}--\cite{marsseno}). The general form of 
a hypersurface that clearly adjusts itself to the spherical symmetry of any 
of these spacetimes is as follows 
\beq 
\Sigma : \cases{\theta = \lambda_{\theta}, \cr 
\varphi = \lambda_{\varphi} , \cr r = r(\lambda) , \cr t = 
t(\lambda) ,} \eeq where $ \{\lambda,\lambda_{\theta}, 
\lambda_{\varphi} \} $ are the parameters of the hypersurface. 
One must thus identify both hypersurfaces in some way.  The 
identification of $ (\lambda_i)_1 $ with $ (\lambda_i)_2 $ (1 and 
2 label each of the spacetimes) is the most natural one, due to 
the symmetry of the above scheme. In the sequel, 1 labels the exterior 
spacetime and 2 the interior one. 
 
In order to match the exterior solution with the interior one, one basically 
demands the coincidence of the first and second fundamental forms of $ 
\Sigma $ at each spacetime ---the other way is to accept the presence of 
{\em singular} mass shells, which would not require the coincidence of the 
second fundamental form, but we will dismiss such unphysical option. In 
order to include the possibility of matching the interior and the exterior 
at null hypersurfaces (e.g. at an event horizon) one can follow the 
formalism in~\cite{marsseno}. It is worth recalling that the exterior region 
is described by a member of the GNRSS family. Thus, we have to consider two 
possibilities: matching a generalized Nariai metric with a GNRSS one and two 
GNRSS metrics with each other. 
\subsection{The junction of a generalized Nariai metric and a GNRSS one} 
In this case one easily gets $ r_1(\lambda) = \gamma = $ const., $ t_1 
(\lambda) = $ const. ---see \cite{tesis8} for full details. This result 
tells us that the junction between a generalized Nariai spacetime and a 
member of the GNRSS family is impossible. It would only happen for a 
(two-dimensional) surface. Therefore, {\em any} member of the Nariai class 
{\em cannot} be regarded as a good candidate in order to represent the 
interior structure of a regular, static BH. 
\subsection{The junction of two GNRSS spacetimes} 
\label{ss-tjo} 
In this case one gets that two members of the GNRSS family match with 
each other if and only if ---see e.g.~\cite{tesis8}--- either $ r_1(\lambda) 
= r_2(\lambda) = R =$ const., $ \dot t_1 = \dot t_2 $, $ [ H ] = [ H' ] = 0 
  $ or $ r_1 + t_1 = r_2 + t_2 = $ const. The last condition, however, 
describes the motion of a null hypersurface and is not an acceptable 
solution in order to describe the {\em matter} inside a static BH. 
Therefore, we reach the conclusion that: {\em The only acceptable 
hypersurfaces fulfilling the matching conditions, that preserve the 
spherical symmetry, between two spacetimes of the GNRSS family, 
are those satisfying $ r_1(\lambda) = r_2(\lambda) = R =$ const., 
$ \dot t_1 = \dot t_2 $, $ [ H ] = [ H' ] = 0  $.} 
 
Without losing generality, one can choose $ t_1  = t_2 = 
\lambda $, because of the global existence of the Killing vector $ 
\partial_t $. Moreover, we realize that the chosen coordinates are 
privileged ones, in which the matching is explicitly $ C^1 $. The 
hypersurface, $ \Sigma $, will be timelike, null or spacelike 
according to $ H < 1$, $ H = 1 $ or $ H > 1 $, respectively. 
 
To summarize, if vacuum polarization is to be the dominant quantum effect, 
the most simple way to construct a regular BH is to build 
it upon GNRSS spacetimes. 
 
\section{Regular interiors of the GNRSS type} 
\label{s-ri} 
As mentioned elsewhere, the exterior region can appropriately be 
characterized by a member of the Kotller-Trefftz class, which is a subclass 
of the GNRSS family. Then, the matching conditions between the exterior and 
the interior regions are: 
\beq 
H_2(R) = H_1( R) = {2m\over R} -{q^2 \over R^2} + {\le \over 3} R^2, \\ 
H'_2(R) = H'_1(R) = -{2 m \over R^2} + {2 q^2 \over R^3} +  {2\le \over 3} R. 
\eeq 
Moreover, the aim here is to focus on those interior solutions which are 
everywhere regular. From the expressions of the Riemann tensor and 
the metric, one sees that this may only be accomplished if 
\beq 
\label{eqf} H_2(0) = 0, \quad H'_2(0) = 0 . 
\eeq 
Thus, we finally encounter four conditions in order to have a 
regular interior solution. 
 
From now on, we will consider $ H_2 $ to be a smooth function 
of the variable $ \tr \equiv r / R $, a most natural hypothesis in view of 
the regular character prescribed for the interior 
solution. In this case, the origin conditions tell us that \beq H_2 (\tr ) = 
\sum_{n=2}^{\infty} {b}_n {\tr}^n . \eeq Now, one has to impose 
the two other conditions. Obviously it is the same to consider $ 
H_2 (\tr ) $ or $ H_2 (\tr-1) $ in the whole procedure. However, 
we will first work with $ H_2 (\tr- 1) $ in order to implement the 
junction conditions directly. From the preceding 
result, one immediately has \beq H_2 = \sum_{n=0}^{\infty} {a}_n 
(\tr-1)^n, \eeq and the junction conditions tell us that \beq 
{a}_0 = H_1(1) = {2m\over R} - {q^2 \over R^2}+ {\le \over 3} R^2, \\ 
{a}_1 = {\dot H}_1(1) = 2\biggl(-{ m \over R} +{q^2 \over R^2}+  {\le \over 
3} R^2 \biggr), 
\eeq 
where $ H_1 (\tr ) = (2m/R)(1/\tr) -(q^2/R^2)(1/\tr^2) + (\le R^2 / 3) \tr^2 
$ and a dot denotes derivation with respect to $ \tr $. 
The following step is to impose regularity of the solution, 
Eqs.~\rf{eqf}. We get 
\beq 
\sum_{n=0}^{\infty}(-1)^n {a}_n = 0, \quad 
\sum_{n=1}^{\infty}(-1)^n n {a}_n = 0, 
\eeq 
which, by virtue of the matching conditions, yield 
\beq 
\sum_{n=0}^{\infty}(-1)^n {a}_{n+2} =  -{4m \over R} + {3 q^2 \over R^2} + 
{\le \over 3}R^2, \\ 
\sum_{n=0}^{\infty}(-1)^n (n+2) {a}_{n+2} = 2 \biggl(-{m \over R} + {q^2 
\over R^2} + {\le \over 3}R^2  \biggr) . 
\eeq 
It is clear that there are infinitely many possible candidates for 
these interiors. 
\subsection{Isotropization} 
\label{s-i} 
Let us further analyze how they behave near the origin. 
Taking into account the expression of $ H_2 $ in powers of $ \tr 
$ and using Eqs.~\rf{eqe}, we get 
\beq 
G_{11} =-{1\over R^2} \sum_{l=2}^{\infty} (l +1) 
{b}_l {\tr}^{l-2}, \quad G_{22} =-{1\over R^2} 
\sum_{l=2}^{\infty} {l +1\choose 2}{b}_l {\tr}^{l-2}. \eeq It is 
then clear that $ G_{11} $ and $ G_{22} $ are different from each 
other. Yet we have the very relevant property that, for {\em any} of these 
spacetimes, it holds \beq \lim_{\tr \to 0} G_{11} = \lim_{\tr \to 0}( 
-G_{00}) = \lim_{\tr \to 0} G_{22} = \lim_{\tr \to 0} G_{33} = -{3 {b}_2 
\over R^2} . 
\eeq 
Whence, we see that a general isotropization of the Einstein tensor ---and 
consequently of the energy-momentum one--- {\it independent of the model} is 
actually accomplished. In terms of $ a_l $ we get \beq 
G_{11} =-{1\over R^2} \sum_{M=0}^{\infty} {A}_M {\tr}^M, \\ 
{A}_M = (-1)^M (M+3) \sum_{l=M+2}^{\infty} (-1)^l {l \choose l-2-M}a_l , \nn 
\eeq 
and 
\beq G_{22} =-{1\over R^2} \sum_{M = 0}^{\infty}{M+2 \over 2} {A}_M {\tr}^M. 
\eeq 
So that 
$$\displaylines{ 
\lim_{\tr \to 0} G_{11} = \lim_{\tr \to 0}( -G_{00}) = \lim_{\tr \to 0} 
G_{22} = \lim_{\tr \to 0} G_{33} = -{ {A}_0 \over R^2}, \cr 
{A}_0 = 3 \sum_{l=2}^{\infty}  (-1)^l{l \choose l-2}{a}_l  . }$$ 
 
Finally, making no further assumptions on the coefficients of 
$ H_2 $, we can isolate two of them in terms of the rest. For simplicity, 
we shall isolate $ a_2 $ and $ a_3 $. The result is 
$$\displaylines{ 
a_2 = -{10 m \over R } + {7 q^2 \over R^2} +{\le \over 3}R^2 + 
\sum_{l=4}^{\infty} (-1)^l(l-3) 
a_l, \cr 
a_3 = -{6 m \over R } +{4 q^2 \over R^2} + \sum_{l=4}^{\infty} (-1)^l(l-2) 
a_l .} $$ 
With this in hand we can write, respectively, the previous expression for 
the central value in terms of $ a_l $ or $ b_l $, $ l \ge 4 $, 
\beq 
\begin{array}{rcl} 
G_{11} (0) = G_{22} (0) & = & \displaystyle - {3 \over R^2} \Biggl[ {8 m 
\over R} - {5 q^2 \over R^2} + {\le \over 3} R^2 + \sum_{l=4}^{\infty} 
(-1)^l {(l-3)(l-2) \over 2} a_l \Biggr], \\ 
& = & \displaystyle- {3 \over R^2} \Biggl[ {8 m \over R} + 
{\le \over 3} R^2 + \sum_{l=4}^{\infty} (l-3) b_l \Biggr]. \cr 
\end{array} \nn 
\eeq 
\section{Examples} 
\label{s-exa} 
We will consider six examples. Two constitute the 
well-known proposals of \cite{ip,fmm,bp} and \cite{dymni,ds}. Two more 
come from a proposal  in \cite{ayon}--\cite{ayon3}, for 
electrically charged bodies, and the proposal given in \cite{bronnikov}, for 
magnetically charged ones. More specifically, we will here derive 
their corresponding analogues, 
within the present scheme (what is actually more than simply re-writing 
those cases).  The remaining two examples 
constitute a family of brand new candidates, which naturally arise from the 
preceding expressions. We will start with this last pair. 
\subsection{Two arbitrary powers} 
\label{ss-tap} 
Let us just make the choice that only two specific powers of $ H(r) 
$, say $ M$ and $ N $, be present. In order to fulfill 
the regularity conditions, both must satisfy $ M $, $ N \ge 2 $. 
However, if we wish to obtain a de Sitter-like behavior at, and near, 
the origin, we 
must necessarily impose that one ---and only one--- of them, say $ M $, 
 be equal to 2. Thus, 
$ H_2(\tr) $ reads $ H_2(\tr) = b_2 \tr^2 + b_N \tr^N $, for  $ N > 2 $, 
with, 
\beq 
b_2 = {2m \over R}\Biggl({N+1 \over N-2} \Biggr) - {q^2 \over 
R^2}\Biggl({N+2 \over N-2} \Biggr) + {\le R^2 \over 3}, 
\quad b_N = {2 \over (N-2) R}\Biggl(-3m + {2q^2 \over R}\Biggr). 
\nn 
\eeq $ 
G_{11}(\tr) $ and $ G_{22}(\tr) $ read ---recall Eqs.~\rf{eqe}--- 
\beq 
G_{11}(\tr) = -\le + 
{6m \over R^3} \Biggl({N+1 \over N-2} \Biggr) \bigl(\tr^{N-2} - 
1 \bigr) + {q^2 \over R^4} \Biggl[{3(N+2) - 4 (N+1) \tr^{N+2} \over N-2} 
\Biggr], \nn\\ 
G_{22}(\tr) = -\le + {6m \over R^3} \Biggl({N+1\over N-2} \Biggr) \Biggl({N 
\over 2}\tr^{N-2} - 1 \Biggr) + {q^2 \over R^4} \Biggl[{3(N+2) - 2 N (N+1) 
\tr^{N+2} \over N-2} \Biggr]. \nn 
\eeq 
Whence, one readily sees that their finite value at the origin 
coincides, as expected, 
\beq 
\label{eqg0} 
G_{11} (0) = G_{22} (0) = -\le - {6m \over  R^3} \Biggl({N+1\over 
N-2}\Biggr) + {3 q^2 \over R^4} \Biggl({N+2\over N-2}\Biggr) . 
\eeq 
\subsubsection{Lowest powers} 
\label{sss-lp} 
This example corresponds to the case in which $ H_2 $ is a polynomial 
of lowest degree. This amounts to setting $ a_l = 0 $, $ l \ge 4 $, in the 
general expressions. Its interest lies in its being the simplest possible 
situation. The result is 
\beq 
H_2(\tr) = \Biggl({8 m \over R} - {5 q^2 \over R^2} + {\le R^2 \over 3} 
\Biggr) \tr^2 + \Biggl(-{6 m \over R} + {4 q^2 \over R^2} \Biggr) \tr^3, 
\eeq 
and 
\beq 
G_{11} = -{24 m \over R^3} + {15q^2 \over R^4} - \le + 
{8 \over R^3} \Biggl( 3m - {2 q^2 \over R}\Biggr) \tr, \\ 
G_{22} =  - {24 m \over r^3} + {15 q^2 \over R^4} - \le 
+ {12 \over R^3} \Biggl( 3m - {2 q^2 \over R} \Biggr) \tr. \eeq 
Notice that $ G_{11} $ tends to $ - \le - q^2/ R^4 $ as $ \tr $ tends to 1, 
the same value as $ G_{11}^{\rm ext.} (\tr = 1) $, in accordance with 
Israel's conditions \cite{israel2}. 
\subsection{Israel and Poisson's model} 
\label{ss-tao} 
In reference \cite{ip} ---see also \cite{fmm}--- a plausible candidate for 
the energy-matter content of the interiors of regular {\em non-charged} BHs 
was proposed. The authors proposed that a singular layer of 
non-inflation\-ary material should exist between the de Sitter core and the 
external Schwarzschild metric. However the usual spirit of matching a 
stellar interior with a vacuum exterior was lost, the reason being 
the unavoidable presence of a singular layer acting as a matter surface 
density. Indeed, in~\cite{bp} it was argued that their approach could be 
improved by imposing a smooth transition from the hypersurface to the de 
Sitter core. Yet this step was not implemented. In any case, it was the only 
available candidate to continue the studies of quantum regular BHs at that 
time. 
The task here will be to see whether this geometrical and physical 
model can be recovered from our analysis. 
 
In order to to do that, we search for a solution within our family which 
be as close as possible 
to this particular solution. What amounts to looking for a de 
Sitter core for small values of $ \tr $ and a quantum contribution of the 
type of {\em the square of the characteristic curvature of Schwarzschild 
spacetime} near the matching hypersurface. 
These features taken into account, we set for the interior 
\beq 
\label{eqi} 
(G_{00})_{\rm int} = -(G_{11})_{\rm int} = {1 \over (A + B r^3)^2}, 
\eeq 
where $ A $ and $ B $ are two constants, to be determined. Using 
Eqs.~\rf{eqe}, we obtain 
\beq 
H_2(r) = {r^2 \over 3 A (A+ B r^3)} , 
\eeq 
where we have imposed Eqs.~\rf{eqf}. The matching conditions lead to 
\beq 
\label{eqh} 
{1 \over A(A+Br^3) }= {3 \over R^2} H_1(R), \quad 
{2A- B R^3 \over A (A+ B R^3)^2} = {3 \over R}H_1'(R), 
\eeq 
where $ H_1 $ comes, as usual, from the external model. 
 
In the exterior region, close to the matching hypersurface, the quantum 
contributions do not turn into a cosmological-like term. They 
 are of the form $ 
(G_{00})_{\rm ext.} \allowbreak \propto m^2/r^6 $, as mentioned before. We 
thus have a quantum exterior which is different 
from the one encountered in the rest of the examples and sections 
before, which cannot be described by a member of the Kottler-Trefftz class. 
Fortunately, our preceding results are still useful. In fact, one realizes that 
it is possible to select a suitable exterior with a similar form 
as~\rf{eqi}, just by setting $ A_{\rm ext.} = 0 $, and $ B_{\rm ext.}^{-1} = 
\alpha m $, where $ \alpha = \beta L_{\rm Pl} $, being $ \beta $ of order 
unity, and $ L_{\rm Pl} $ the Planck length ($\alpha$ is of order 
unity in Planckian units).  $ \beta^2 $ is related with the number 
and type of the quantized fields, \cite{ip,bp}. This choice 
yields 
\beq 
\label{eqj} 
H_1(r) = {2 m \over r}- {1 \over 3}\biggl( {\alpha m \over  r^2}\biggr)^2, 
\eeq 
where we have taken into account that the exterior region is dominated by 
the Schwarz\-schild geometry ---with mass $m $--- for large values of $ \tr 
$. 
 
Now, using Eqs.~\rf{eqh} and~\rf{eqj}, $ A $ and $ B $  yield 
\beq 
A = {\alpha \over 6- \displaystyle{\alpha^2 m \over R^3}}, \quad 
B = {2 \over \alpha m}\left(1-{ 3  \over 6 - \displaystyle{\alpha^2 m \over 
R^3} }\right). 
\eeq 
Finally, using Eqs.~\rf{eqi}, $ (G_{22})_{\rm int} = (2B r^3 - A)/(A + B 
r^3)^2 $, where $ A $, $ B $ have been given above. At the origin 
\beq 
\label{eqm} 
G_{11}(0) = G_{22} (0) = -{1 \over B^2}. 
\eeq 
 
To summarize, we have proven here that a spacetime model within our family 
satisfies all the required geometrical assumptions, and yields the particular 
form of $ G_{00} $, both for the interior and the exterior of the body, as 
 in the above mentioned references. A more throughout comparison of 
that model and ours will be given in Sect.~\ref{ss-tao2}. 
\subsection{Dymnikova's model} 
\label{ss-dym} 
Some time after the appearance of the previous cases a new model for a 
regular interior of a {\em non-charged} BH was proposed \cite{dymni}. 
However the approach was now quite different to that of the previous 
authors. Now Schwarzschild's solution was only recovered in an asymptotical 
sense, for $ \tr $ approaching infinity only. However, if 
a sufficiently quick convergent matter model was obtained, then the quantity 
of mass outside the horizon of the collapsed body could become as negligible 
as desired with regard to the interior mass. Thus one would, at least, 
recover a trial model, interesting enough to support or reject the 
conclusions of the previous authors. In a later work \cite{ds}, the model 
was extended to incorporate the observational fact in favor of a 
non-vanishing cosmological term in the exterior region. We will deal in 
this subsection with such model, but considering a definite end to the 
collapsed body. 
 
The imposition for the energy-matter content for the interior will be of the 
form 
\beq 
(G_{00})_{\rm int} = -(G_{11})_{\rm int} = A \exp{(- \tr^3)} + B, 
\eeq 
where $ A $ and $ B $ are two constants to be determined and $ \tr \equiv r/ 
R $, where $ R $ is the matching radius.\footnote{For other choices see e.g. 
\cite{tesis8}.} We then integrate the expression of $ G_{00} $ ---recall 
Eqs.~\rf{eqe}--- in order to obtain $ H_2 $, getting \beq 
H_2(\tr) = {R^2 \over 3} \Biggl[{A \over \tr}\Bigl(1-e^{-\tr^3}\Bigr) + B 
\tr^2 \Biggr], 
\eeq 
where we have already imposed the regularity conditions at the origin, 
Eqs.~\rf{eqf}. The matching conditions at the spatial hypersurface yield 
\beq 
A {(e-1) \over e} + B =  {6m\over R^3} + \le \nn \\ 
A {(4-e) \over e} + 2B = -{6m\over R^3} + 2 \le\nn 
\eeq 
whence, 
\beq 
A = {6m \over R^3} \Biggl({e \over e-2} \Biggr), \quad B= \le - {6m \over 
R^3} {1 \over (e-2)} . 
\eeq 
Finally, using Eqs.~\rf{eqi}, 
\beq 
(G_{22})_{\rm int} (\tr) = A \Bigl( {3 \over 2} \tr^3 - 1 \Bigr) e^{\-\tr^3} 
- B, \nn 
\eeq 
where $ A $ and $ B $ have been given above. At the origin 
\beq 
\label{eqt} 
G_{11} (0) = G_{22} (0) = \le + {6m \over R^3} \Biggl( {e-1 \over e-2 } 
\Biggr) . 
\eeq 
In Sect.~\ref{ss-tap2} we will compare, numerically, our results with those 
in the model of \cite{dymni}. 
\subsection{Ay\'on--Beato and Garc\'\i a's models} 
\label{ss-abg} 
In a series of papers, \cite{ayon}--\cite{ayon3}, some models of regular, 
{\em electrically} charged BHs with an energy-momentum tensor of the form of 
a SSQV were presented ---see also \cite{bardeen}. Their importance relied in 
the fact that the sources that give rise to those spacetimes could be linked 
with non-linear electrodynamics (NED), which besides being a theory by 
itself, may be viewed as a low energy limit of string theory or M-theory. 
Thus, some plausible models of regular BHs ---that took into account quantum 
effects in a clearer way than before--- were put forward.  The features of 
their models are analogous to the case of Dymnikova's model, though with a 
clear interpretation of the source origin. For the sake of brevity, we will 
focus on the model in \cite{ayon3}. 
 
The choice there was $ H(r) = (2m/r)[1-\tanh{(q^2/2mr)}] $, for 
any $ r \geq 0 $.  Ours will be: 
\beq 
H(r) = \cases{ \displaystyle {A \over r} \Biggl[ 1- \tanh{B \over r}\Biggr], 
& $ 0 \leq r \leq R $, \cr \displaystyle {2m \over r} - {q^2 \over r^2}, & $ 
R \leq r $, \cr} 
\eeq 
where $ A $ and $ B $ are constants to be determined. The matching 
conditions imply 
\beq 
A = {q^2 \over B} \cosh^2{B \over r}, 
\quad 1 + e^{\textstyle -{2 B \over R }} = 2B \biggl( {2m \over q^2} - {1 
\over R} \biggr). 
\eeq 
Defining $ x \equiv A/ A_0 $, $ y \equiv  B/ B_0 $ with $ A_0 = 2m $, $ B_0 
= q^2/2m $, that is the values of the model in \cite{ayon3}, we get 
\beq 
\label{eqxy} 
x= {1 \over y} \cosh^2{\lambda y}, \quad 1 + e^{\textstyle - \lambda y} = 
2(1- \lambda) y, 
\eeq 
where $ \lambda \equiv B_0/ R = q^2/2mR $. One has here to solve a 
trascendental equation in order to find the appropriate constants of the 
interior model. The parameter $ \lambda $ is  the one controlling the 
set of solutions. In classical electrodynamics, $ \lambda = 1$. We see 
that there is no solution in this case. In the context of General 
Relativity, $ \lambda = 1 $ corresponds to the case where the exterior 
metric becomes flat at a spherical surface. But the choice of $ H_{\rm int} 
$ cannot be zero for any positive value of $ r $. Therefore the matching is 
impossible. The same happens for the other models  in 
\cite{ayon}--\cite{bardeen}. In the following section we will see which 
type of solutions arise for different values of $ \lambda $. 
\subsection{Bronnikov's model} 
\label{ss-bm} 
In \cite{bronnikov} a model for static, regular, {\em purely} magnetically 
charged BHs with an energy-momentum tensor of the type of SSQVs was 
proposed. Its interest is two-fold. Again the energy-momentum content of the 
objects was directly connected with NED. Second, it turns out that those BHs 
are the only ones based on a Lagrangian formulation of NED with a Maxwellian 
behaviour in the weak field limit, regardless of the place the weak limit  is 
taken. The example given there was a GNRSS metric with $ H(r) = 
(|q_m|^{3/2}/ a r)[1-\tanh(a\sqrt{|q_m|}/r)] $ with $ m $ ---the ADM mass--- 
equal to $ |q_m|^{3/2}/2a $, being $ q_m $  the magnetic charge. It is then 
obvious that the results of the previous subsection are valid now, just by 
changing $ q $ with $ q_m $. The difference lies in the fact that now one has 
magnetic fields and also the theory describing NED is different to that of 
\cite{ayon}--\cite{ayon3}. 
\section{Numerical results} 
In order to study the approximate values of $ R $ for a given object, one 
needs to assume a particular behaviour of the matter and energy inside the 
source. As of now, there is no agreement at this point. However, following 
several results, see e.g. \cite{glinner}--\cite{ayon}, 
\cite{burinskii}--\cite{cvs}, the geometry of the core may be described 
by a dS solution. This  has been the assumption used in most of the 
 works dealing 
with regularized BHs. Here we will also include two examples with a 
different behaviour and in Sect.~\ref{s-tmo} we will draw the main lines of 
a general behaviour. In any case, the aim is to choose those physical models 
which are as consistent as possible, the dS model being one of them. 
In this case, 
at the core we will have $ G_{00}(0) = - G_{11}(0) = - G_{22}(0) = 
-G_{33}(0) =  \li = {\rm const.} $ 
Nonetheless, there is no present agreement about the scale at which 
regularization could act. A convenient way to handle and integrate this 
indeterminacy is to set $ \li = 10^{3s}\le $, being $ s $ the free 
parameter that governs the renormalization scale. For instance, if $ s $ 
is around 40, we are then considering that regularization takes place at 
Planck scales, and so on. 
 
Finally, for the exterior region, in accordance to several recent 
observations \cite{perl,ries}, we will assume  in what follows that $ 
\rho_{\le}\! \in \! [10^{-10}, 2 \times 10^{-8}] \;{\rm erg 
\!\cdot\! cm^{-3}} $. An analysis shows, however, that the fundamental 
contribution comes from the quantum gravitational model describing 
the core, and not from the type of quantum vacuum contribution that is 
assumed for the exterior region or near the surface of the body. 
 
\subsection{Two arbitrary powers} 
\label{ss-tap2} 
In this numerical analysis we will consider the uncharged case, because 
there are no observed objects that can be associated with static, charged 
BHs. If there is charge in the source, then the interesting situation 
involves 
rotation, which might be eventually connected with elementary 
particles (we refer the reader to~\cite{behm}). The 
relation~\rf{eqg0} is (now $ q = 0 $) 
\beq 
\le = \li + {6m \over R^3} 
\Biggl({N + 1 \over N - 2} \Biggr), \quad \forall N \ge 3, 
\eeq 
whence 
\beq 
\label{eqz} 
R = R_{\odot} \root{3}\of{M} \root{3}\of{N+1 \over 4(N-2)}, 
\eeq 
where we have put $ m = M m_{\odot} $, $ m_{\odot} $ being 
the Sun mass, and $ R_{s} \equiv $ $ \root{3}\of{24m_{\odot}/ 
}\allowbreak\overline{(\li -\le) }$. The last value only depends on the 
regularization scale and corresponds to the solution for a collapsed 
object of one solar mass in the case of the ``lowest powers'' model: $ R_{s} 
\in [3 \times 10^{21-s}, 6 \times 10^{20-s}] {\rm cm} $. For $ s= 40 $ we 
get $ R_{s} \in [3 \times 10^{-19}, 2 \times 10^{-20}] {\rm cm} $. Yet we 
see that the object has a quantum size very far from 
Planckian scales, even if $ s $ is bigger. In general $ R_{s} / 
L_{\rm Pl} \ge 10^{13}!$ Moreover this result is valid for all $ N 
$, since for any value of $ N $ we have that $ R \in [0.6  , 1] R_{s} 
\root{3}\of{M} $. It is obvious that, for any astrophysical object, 
the final properties are very similar. Table~\ref{tab-1} comprises 
different massive objects and regularization scales and their associated 
values of $ R $ within this model. 
\begin{table} 
$$ 
\begin{array}{||c||c|c|c|} 
\hline\hline 
m & s= 30 & s= 40 & s= 50 \cr 
\hline 
\hline 
M_{\odot}  & 10^{-9}& 10^{-19} & 10^{-29} \cr 
\hline 
10^3 \,  M_{\odot} & 10^{-8}& 10^{-18} & 10^{-28} \cr 
\hline 
10^6 \, M_{\odot} & 10^{-7}& 10^{-17} & 10^{-27} \cr 
\hline 
10^9 \, M_{\odot} & 10^{-6}& 10^{-16} & 10^{-26} \cr 
\hline 
\end{array} $$ 
\caption{\label{tab-1} {\protect\small 
$ R $ in cm for various astrophysical and galactic 
objects and different scales of regularization ($ s = 30 $ corresponds 
to a GUT's regularization scale, $ s = 40 $ to a Planckian one, etc.). In 
any case $ R/L_{\rm Reg} $ is much bigger than 1 ($ R/L_{\rm Reg} 
\sim 10^{(-6 + s/2)}$). Therefore, all of them are quite far from their 
corresponding regularization scale.}} 
\end{table} 
\subsection{Israel and Poisson's model} 
\label{ss-tao2} 
We have found that the corresponding model within our family 
must satisfy 
\beq 
A = {\alpha \over 6- \displaystyle{\alpha^2 m \over R^3}}. 
\eeq 
In this case, $ A^{-2} = \lim_{\tr \to 0} G_{00} = \li $, so that 
\beq 
R^3 = {\alpha^2 m \over 6 - \alpha\sqrt{\li}}={\beta^2 \over 6 
- \beta \sqrt{\li L_{Pl}^2}} m L_{\rm Pl}^2 . \eeq This model 
clearly depends on the coefficient $ \beta $. For instance, in 
order to obtain a solution, we must have $  \beta^2 < 36/(\li 
L_{\rm Pl}^2) $. The natural scale of regularization in this model 
is the Planckian one since from the beginning the coefficient $ 
\alpha $ was related to the Planck length. Obviously other 
regularization scales would simply change $ L_{\rm Pl} $ by the 
corresponding scale. Using standard values for $ \li $, that use 
a Planckian regularization scale, and the fact that $ \beta^2 $ should 
be at most of order unity \cite{ip,bp}, we get $ R \sim 
\root{3}\of{M} \times 10^{-20} {\rm cm.} $ This result is in 
complete agreement with the foregoing values, even though the 
models possess very different functions $ H (r) $. 
\subsection{Dymnikova's model} 
\label{ss-dm2} 
From Sect.~\ref{ss-dym} and the assumption of a dS core, we have 
\beq 
\le + {6 m \over R^3} \Biggl( {e-1 \over e-2 } \Biggr) = \li, 
\eeq 
whence 
\beq 
R = \root{3}\of{{6m \over \li - \le}\left({e-1 \over e-2}\right)}. 
\eeq 
Comparing this result with the one in Eq.~\rf{eqz}, we get that $ R = \lambda 
R_{\rm Two powers} $, with $.84 < \lambda < 1.34 $, for any N. Therefore, $ 
R $ is again of the same order of magnitude, despite the differences in the 
choice of the profile of the energy density and of the tangential pressures. 
 
Comparing now the model proposed here with the original one in \cite{dymni}, 
we see that both yield similar conclusions (in the instances they can be 
compared). For example, in the mentioned work, 
 a characteristic radius was found for 
the collapsed body. Its expression is $ R_c = \root{3}\of{6m/(\li -\le)} $, 
what yields $ \root{3}\of{(e-2)/(e-1)} R \sim 1.34 R $. Besides 
 having a different description for this in our model, the values of the 
coefficients $ A $ and $ B $ are also quite different (numerically). 
\subsection{Ay\'on--Beato and Garc\'\i a's model, and Bronnikov's model} 
\label{ss-abg2} 
In the papers dealing with those models, there is no analysis of the orders 
of magnitude of an eventual characteristic radius. The only such condition on 
 these 
model is to 
have an event horizon. We can now compute which are the ranges of $ R $ 
corresponding to different cases of $ \lambda $. 
 
First of all, Eqs.~\rf{eqxy} only have solution for  $ 0 < \lambda < 1 $. 
Therefore, extremely charged objects (those with $ |q|/m >> 1 $), cannot be 
described within the present framework. This would require $ R >> m $, so that 
the regularized object would not be a BH but a ``visible" object, such as an 
electron (its size, though, being bigger than the classical radius, or 
Compton size, $ q^2/ 2m $). 
 
For strongly charged objects, i.e. $ |q|/ m \sim 1 $, we get that, in order 
to have a BH, $ {1}/{2} < {R}/{m} < 2 $. Thus, the regularized object is of 
a similar size as that of the event horizon. Much bigger than in the uncharged 
case. 
 
The solution given by \cite{ayon3}, i.e. $ A = 2m $, $ B = q^2/ 2m $, 
 can 
only be valid now for very weakly charged objects, $ |q|/m << 1$, and 
satisfying $ R/m < q^2/m^2 $. They showed that their model was acceptable 
for $ |q| /m \leq 1.05 $. Now, we see that the values of $ A $, $ B $ in our 
model change for most of these cases. 
 
The same is valid for Bronnikov's model, just changing the electric 
field by a magnetic one. Nevertheless, rotation should be introduced in 
such case ---at least when $ |q|/m $ is not very small. 
\section{Horizons and an interpretation of the regularized BH} 
\label{s-hirbh} 
Looking at Eq. (30) in \cite{ds} and  comparing it with our result 
\beq 
g_{00} = -1 + H_2 , 
\eeq we realize that, substituting here our corresponding $ H_2 $ 
for that model, these expressions turn out to be very similar, except 
for a possible overall sign difference, due to the different signatures (e.g. 
(+,--,--,--) instead of our (--,+,+,+)). We conclude that the same 
structure for the horizons and Cauchy hypersurfaces is obtained. 
In \cite{ds}, the solutions are obtained by approximations of the 
exact solution, so that these results and ours are really 
coincident (the relative error  with 
respect to both exact solutions being completely negligible). 
 
In general, the horizons result from the cancellation of $ g_{00} 
$. Thus we are left with a general set of horizons. A global 
study for all the candidates encountered, has not yet been carried 
out. We could focus on examples, and try then to extract some 
general features from them, but we do not find this of primary 
importance.\footnote{With respect to the other models encountered here, 
we have found that the results are rather similar to those in 
 Dymnikova and Soltysek's model 
\cite{ds}.} The main point is here, in fact, that the matching 
occurs at a radius which is substantially smaller than the 
Schwarzschild radius of the object. Therefore we will always have 
a typical exterior, a vacuum transition region extending until the 
matching with the object happens, and a quantum-dominated 
interior, which finally converges to a de Sitter core. In the 
vacuum interior region and in some part of the quantum object, the 
role of $ t $ and $ r $ are not interpreted as usual ($\partial_t 
$ changes its character). This is the reason for adequately 
treating the horizons: to see where exactly such changes appear. 
But, we can still perfectly agree in ordinary physical terms 
without requiring a general determination of the precise radii at 
which horizons occur. 
 
Moreover, in \cite{bp}, the authors studied the stability of the model. 
The same considerations there  hold 
for our whole family of solutions, as can be easily seen after a careful 
analysis. 
 
Finally, there is still the issue of the topology of the solutions, which is 
connected with the possibility of a ``universe reborn''  in the extended 
spacetime. Its general structure can be found in \cite{borde} for the case 
where the sources satisfy weak energy conditions (see next section). There, 
it was shown that the topology of any regular BH, satisfying the weak energy 
conditions,  should be similar to that of a singularity-free 
Reissner-Nordstr\"om spacetime. However, there are relevant solutions in our 
family that violate the weak 
energy conditions (WEC). It would be worth studying what happens in those 
cases. 
\section{Energy conditions} 
\label{s-ecs} 
A common point when dealing with the avoidance of singularities is to show 
that the energy conditions  required in the singularity theorems (see e.g. 
\cite{he}) fail to be valid. 
 
Here we will study the strong energy conditions (SEC), the weak 
energy conditions (WEC), the null energy conditions (NEC) and the dominant 
energy conditions (DEC), within the GNRSS family (see  \cite{mps} for the 
case of a general spherically symmetric spacetime). The SEC are related with 
the  formation of singularities in the collapse of an 
object. The WEC are directly related with the energy density measured 
by an observer. The NEC are useful in order to include some spacetimes which 
violate the first two, but are predicted by some quantum models, e.g. AdS. 
Finally, the DEC are in fact related with the causal structure of the 
energy-matter content of a spacetime \cite{bm}.\footnote{Let us notice 
by passing  that  their Eqs.~(2.25), expressing the DEC, are wrong. 
For our case, the correct ones 
are given in Eqs.~\rf{dec}.} Even though an analysis of energy conditions 
helps to understand the physics of a model, one has to be cautious 
on ascribing 
to them more relevance than they actually have. In several systems, mainly 
when quantum effects play a fundamental role, they all may be violated with 
less difficulty (see e.g. the review in \cite{visser}). 
 
Let $ \{ {\vec e}_a \} $, $a=0,1,2,3$, be a dual vector basis of the cobasis 
used in~\ref{ss-tgn}, defined by $ \th^b {\vec e}_a \equiv \delta^b_a $, $ 
b=0,1,2,3$. Any timelike vector field, $ {\vec V} $, in the manifold can be 
represented by 
\beq 
\label{4vel} 
{\vec V} = A^b {\vec e}_b, \qquad (A^0)^2 = 1 + \sum_{i=1}^{3} (A^i)^2, 
\eeq 
where $ A^b $ are some functions. 
 
On the other hand, from the results of Sect.~\ref{s-swa}, the 
Ricci tensor is \beq {\bf Ricci} = R_{00} (\th^0\otimes\th^0 - 
\th^1 \otimes \th^1) + R_{22}(\th^2\otimes\th^2 + \th^3 \otimes 
\th^3), \eeq 
where $ \otimes $ denotes tensor product. A similar expression holds for the 
Einstein tensor. 
\subsection{Strong energy conditions} 
SEC require $ R_{VV} \equiv R_{ab}V^aV^b \ge 0 $, for all $ \vec V $. From 
the expressions above, we obtain $ R_{VV} = R_{00} + (R_{00} + 
R_{22})[(A^2)^2+(A^3)^2] $. Taking into account the expressions of the Ricci 
and Einstein's tensor given in Sect.~\ref{ss-tgn} we get $ R_{VV} = G_{22} + 
(G_{00} + G_{22})[(A^2)^2+(A^3)^2] $. Finally, using Einstein's equations, 
and the fact that $ A^2 $, $ A^3$ are free, we get 
\beq 
{\rm SEC} \leftrightarrow \rho + p_2 \ge 0, \quad p_2 \ge 0 
\eeq 
where $ \rho $ is the energy density measured by $ {\vec e}_0 $, 
$ 8 \pi \rho = G_{00} $ and $ p_2 $ is the tangential pressure (or stress) 
of the source, $ 8 \pi p_2 = G_{22} $. This is the usual representation of 
SEC. However, the GNRSS family allows for a {\em new}, and more useful, 
expression. Indeed, as mentioned elsewhere, $ p_2 = -(\rho + r\rho'/2) $, 
where $ ()' \equiv d()/dr $. 
Therefore, we can write 
\beq 
{\rm SEC} \leftrightarrow p_2 \ge 0, \quad \rho' \leq 0 . 
\eeq 
 
In all the examples given before SEC are violated. This is natural since 
they are regular. Particularly,  SEC are violated for $ r \leq R_{SEC} $, 
with 
\beq 
R_{SEC} = \root {N-2}\of{2 \over N} R, \ {\textstyle R_{SEC} = \left( 
\root{3}\of{ \alpha^2 m \over 
4(6R^3-\alpha^2m)} \right) R},  \ R_{SEC} = .68 R, \ R_{SEC} = .83 B \nn 
\eeq 
where, all the quantities have been defined in Sect.~\ref{s-exa} and the 
solutions correspond to the two-power model, the Israel-Poison's model, 
Dymnikova's model, and Ay\'on--Garc\'\i a's \cite{ayon3} and Bronnikov's 
model, respectively. Indeed, SEC are violated in a main portion of the 
object, i.e. $ R_{SEC} \stackrel{<}{\sim} R $. For the evaluation of the 
Israel-Poison's model, we have used the same numerical values as in 
Sect.~\ref{ss-tap2}. In the case of Dymnikova's model the value displayed 
corresponds to the case $ \li >> \le $. For any other case with $ \li > \le 
 > 0 $  or $ \li <0 < \le $, as expected, SEC are violated from bigger 
values of $ R_{SEC} $. Finally, in the latter case, one should evaluate $ B 
$ for different possibilities (see Sect.~\ref{ss-abg} and the next case). 
\subsection{Weak energy conditions} 
Following analogous steps, one finds, for WEC ($G_{VV} \geq 0 $, for all $ 
\vec V $) 
\beq 
{\rm WEC} \leftrightarrow \rho \ge 0, \quad \rho' \leq 0 . 
\eeq 
 
It turns out that WEC are satisfied in the models of Sects.~\ref{ss-tap}, 
Sect.~\ref{ss-tao} and Sect.~\ref{ss-dym}, very easily for any value of $ r $ 
(e.g. for de Sitter core, $ \rho'=0 $). One only needs 
to impose $ \le < \li $. 
 
Let us now consider the series of models in Sect.~\ref{ss-abg}, 
~\ref{ss-bm}.  We have already seen that $ 0 < \lambda < 1 $. This implies 
that $ y, x > 0 $, and hence that $ A, B >0 $. In general, we have $ H(r) = 
(A/r) [1-\tanh(B/r)] $. We then get $ 8 \pi \rho = (1/r^2) (Hr)' = 
(AB/r^4) \, \cosh^{-2}(B/r) $ and $ 8 \pi \rho' = (2 A B / r^5) \, 
\cosh^{-2}(B/r) \times [-2 + (B/r) \tanh(B/r)] $. The energy density $ 
\rho $ is positive for any $ r $, although $ \rho' $ may become positive. To 
see this, we first solve $ \rho'  = 0 $. Its solution is $ r \simeq .48 B $. 
Therefore, we have: For $ r < .48 B \cap r < R $ (outside the body WEC are 
satisfied), WEC are violated. 
 
In the model of Refs.~\cite{ayon3},~\cite{bronnikov}, one has $ B = q^2/2m $ 
and $ |q| < 1.05 m $. This gives that WEC are violated for $ r < .27 m $, 
already far away from the core. 
 
In our revisited model, we have basically two different possibilities. First, 
for weakly charged sources, i.e., those with $ |q|/m << 1$, the conclusions 
are the same as for the model in \cite{ayon3},~\cite{bronnikov}. Second, for 
sources with $ |q|/m \sim 1$, we have ---recall Sect.~\ref{ss-abg2}--- $ m/2 
< R < m $, for a BH. 
Two limiting alternatives appear. 
 
The first one is that $ R \to 2m $. In this case $ y \sim 1 $ and, 
therefore, WEC are violated for $ r < .27 m $. The other  one is that $ R 
\to (m/2) + \epsilon $, with $ \epsilon << 1$. Now, $ y \sim 1/4\epsilon $. 
WEC are violated for $ r < .06 m / \epsilon \cap r < 2 m $, that is 
everywhere inside the source. In conclusion, WEC are again violated almost 
everywhere. 
 
Finally, if one lets $ R > 2m $ (one does not have now a BH, but a ``visible'' 
object) big enough to have $ \lambda < 1 $ for any $ |q| $, $ m $, we get 
that WEC are violated everywhere in the object. 
 
This adds a new (elementary) example to the violation of WEC when quantum 
effects play an important role (see \cite{visser} for a recent review) and 
shows clearly that, although energy conditions do help understanding the 
models, they should  not 
necessarily restrict the search for new solutions (Fig. 1). 
 
\begin{figure} 
\centerline{\epsfxsize=14cm \epsfbox{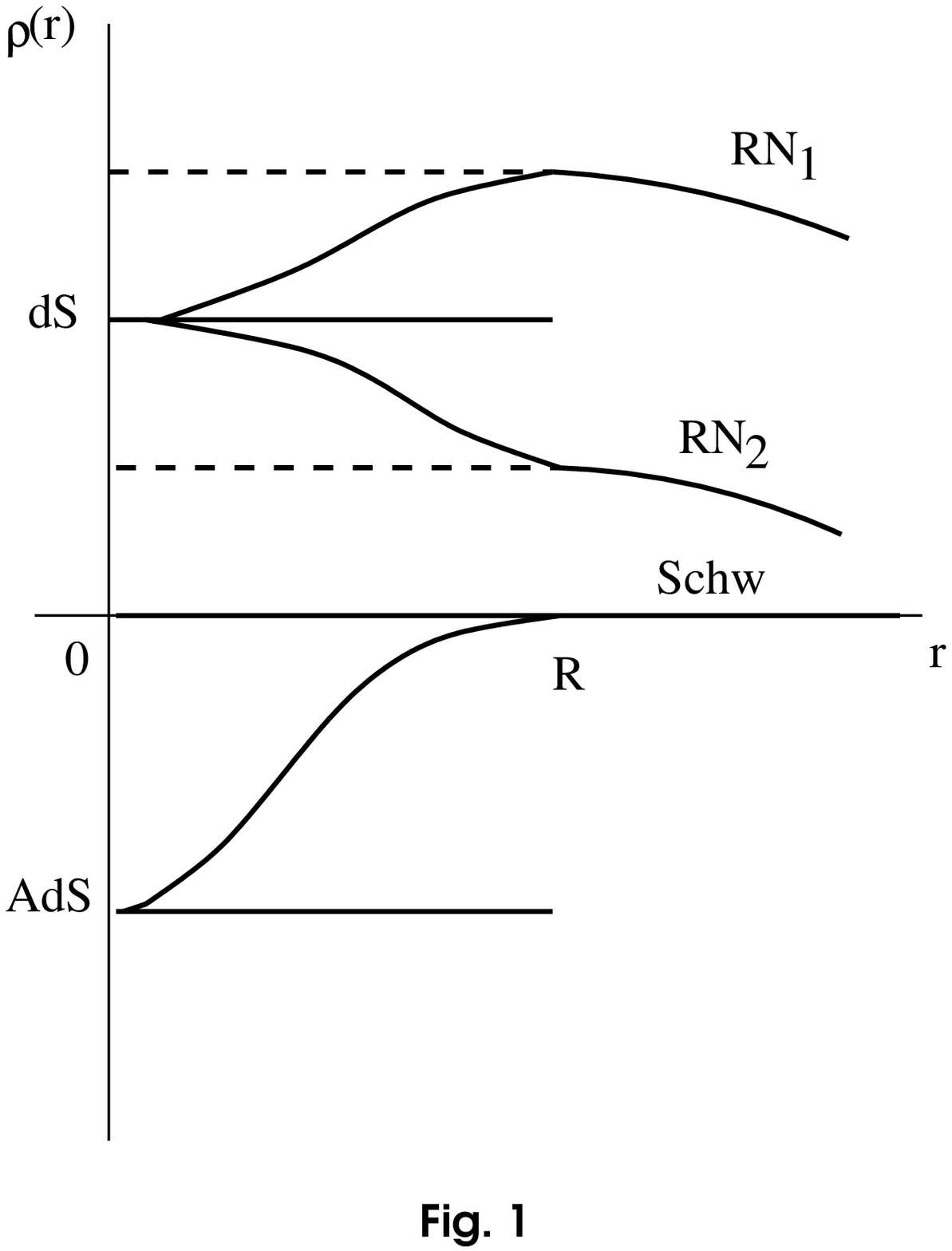}} 
\caption{{\protect\small 
Plot of the density $ \rho $, in arbitrary units, in terms 
of the coordinate $ r $. RN means Reissner-Nordstr\"om, dS means de 
Sitter, Schw means Schwarzschild and AdS means anti de Sitter. $ 
\rho_{RN} = e^2/r^4 $, $ \rho_{dS} = \li $, $ \rho_{Schw} = 0 $ and 
$ \rho_{AdS} = - |\li| $. In the region $ r \in [0,R]$, $ \rho $ can 
be any (smooth) function matching continuously with $ \rho $ at the 
center and at the surface. Regions where $ \rho $ is increasing 
violate SEC, WEC and NEC. These are clearly most but not all 
possibilities. From the plot, $ e^2/R^4 < 8 \pi \li $, if one wants 
that the model fulfills WEC or NEC. The addition of an external $ 
\Lambda $ simply shifts the horizontal axis a quantity $ \le $.}} 
\label{f1} 
\end{figure} 
 
\subsection{Null energy conditions} 
In the case of NEC, $ \vec V $ is a {\em null} vector field, $ \vec V \cdot 
\vec V = 0 $, and requires the evaluation of $ R_{ab} V^a V^b = G_{ab} 
V^aV^b \ge 0, \forall \vec V $. One obtains 
\beq 
{\rm NEC} \leftrightarrow \rho' \leq 0 . 
\eeq 
Thus one sees, that a necessary condition {\em common to SEC, WEC and NEC} 
is that the energy profile of the sources be a non-increasing function. NEC 
are satisfied in the models of Sects.~\ref{ss-tap},~\ref{ss-tao} 
and~\ref{ss-dym}, for $ \li > \le $, regardless of the signs 
 of $ \li $ or $ \le 
$. In the models of Sects.~\ref{ss-abg},~\ref{ss-bm}, NEC are violated in 
the same regions as WEC are, contrary to the belief expressed in 
\cite{bronnikov}. 
\subsection{Dominant energy conditions} 
DEC are satisfied if and only if $ |T^0_0| \ge |T^i_j| $, $ i,j  = 1,2,3 $. 
For the GNRSS family one gets 
\beq 
\label{dec} 
{\rm DEC} \leftrightarrow {\rm sign} {\rho} = {\rm sign}{(- \rho'}) = {\rm 
sign} (\rho' + 4 \rho /r) . 
\eeq 
Two immediate consequences are, that if $ \rho $ changes its sign, DEC are 
violated, and if WEC are violated in a region with $ \rho \ge 0 $, then 
 DEC are  also violated. 
 
Let us turn now to the models considered here. In the case of the {\em 
two-powers} model we will assume $ \le \geq 0 $. In this case, $ \rho $ is 
positive everywhere. WEC were satisfied in these models. However, DEC may be 
violated. A study of the sign of $ \rho' + 4 \rho /r $ tells us that SEC are 
satisfied for $ \tr^{N-2} \leq 4 \li/(\li - \le) \times 1 /(N+2) $. Now the 
question is whether $ \tr $ is less than 1 or not. 
 
Obviously, for any $ N $ exceeding $ N^* \equiv 4 \li / (\li - \le ) - 2 = 2 
(\li + \le )/(\li - \le ) $, we have that DEC are violated. One may ask 
whether this is too odd or easy. Since one expects $ \li >> \le $, we 
readily get $ N^* \simeq 2 $. This, together with the fact that $ N $ must 
be bigger than 2 (in order to be singularity-free), implies that, in practice, 
DEC are violated in these models ---recall WEC are satisfied {\em 
throughout}. 
 
For the model of Sect.~\ref{ss-tao}, we can assume $ B $, $ C $ to be 
positive (for, if $ B C < 0 $ one gets a negative Schwarzschild's mass outside 
the body and if $ B, C < 0 $, $ B, C $ can be substituted by $ |B|, |C| $). 
Following a similar analysis as with the previous models, one gets that DEC 
are satisfied for $ r \leq r^* $, where $ r^* \equiv (2B/C)^{1/3} $. 
Therefore if $ r^* $ is less than the matching $ R $, there is a region 
where DEC are violated. This is the case of our corrected model. 
Incidentally, in the original model, DEC are violated for $ r > r^* $. This 
conclusion is against physics, since far enough one expects Schwarzachild's 
solution to be valid, and it is a vacuum's solution with no problems in its 
causal structure. Therefore, the corrected version not only describes a more 
realistic picture but also solves this undesired property. 
 
Turning back to our corrected model we have to answer whether $ r^* $ may be 
smaller than $ R $ ---see the expressions given in Sect.~\ref{ss-tao}. We 
get that for $ \beta  > \beta_+ $ or $ \beta < \beta_- $, where $ \beta_+ 
\equiv 6(\sqrt{3} -1) \times \sqrt{(L_{\rm Pl}^2 \li)} $ and $ \beta_- 
\equiv  -6(\sqrt{3}+1) \times \sqrt{(L_{\rm Pl}^2 \li)} $. On the other 
hand, it is expected, \cite{bd}, that $ L_{\rm Pl}^2 \li \sim {\cal O}(1) $ 
and, consequently,  $ \beta_+ $, $ \beta_- $ are of order unity. 
Therefore, 
even though there are several parameters for which DEC may hold, there are 
also many others for which DEC will fail. 
A more definite answer can only be provided after a particular field model 
is chosen, what will yield a particular value of $ \beta $. What is this 
plausible field model remains, 
as of now, unknown. 
 
For the next model (the one in Sect.~\ref{ss-dym}), it is easy to show 
that DEC are satisfied throughout the source if $ \li > \le > 0 $, as 
expected. This is contrary to the other models, 
since this one departs from them through 
the causal connection in its stress-energy content. It is to be noticed that 
DEC give a new input to understand the models. (The case $ \li < \le < 0 $ 
also satisfies DEC, whereas the rest of possibilities violates them). 
 
Finally, for the models in Sects.~\ref{ss-abg},~\ref{ss-bm}, as $ \rho $ is 
positive and $ \rho' $ is positive near the core, DEC are violated together 
with WEC. 
 
Some concluding remarks are in order. First, although DEC are known to be 
different from  WEC, here we see more: it turns out that in cases 
with $ \rho \geq 0 $, DEC are {\em more 
restrictive} than WEC. Another consequence is that DEC violation and the 
spacetime region where it occurs are not related. That is, DEC may be 
violated in regions where $ H(r) $ is larger or smaller than 1. It happens, 
however, that after substituting expected numerical values for the physical 
parameters involved, the values of $ H(r) $ where  DEC is violated belong 
 mainly 
in the region where $ H(r) \geq 1 $ and a ``signature change in spacetime has 
occured''. The region with $ H(r) \leq 1 $ is then at Planckian 
(regularization) scales and can thus be forgotten. 
On the other hand, when WEC are violated, one usually accepts that the 
energy-matter content of the model can no longer be described by a classical 
matter source model. However, DEC deserve some especial attention. 
 
These remarks impel us to further interprete the violation of DEC from the 
causal interpretation of DEC \cite{bm}. A possibility is that the breakdown 
of  causality in matter interaction may be interpreted in similar ---though 
properly adapted--- terms as is the Einstein-Podolski-Rosen paradox 
interpreted in Quantum Mechanics. 
 
If this is so, or something similar can be proven, DEC may be a more natural 
sign of quantum effects in matter than WEC, for the case of positive 
densities. This point deserves further investigation. 
 
We will now analyze the main features arising when one replaces the 
 de Sitter core by  a different spherically symmetric solution. 
\section{The matching of static spherically symmetric \allowbreak 
space\-times} 
\label{s-tmo} 
In previous sections we have worked with the assumption that the 
energy-momentum tensor satisfies $ T^0_0 = T^1_1 $. Now we would like to 
make the first steps towards the general case where $ T^0_0 $ and $ T^1_1 $ 
may be independent of each other. Therefore, our aim here is to match two 
spacetimes that share the existence of an integrable Killing field and 
spherical symmetry. In order to get the most natural junction, we need to 
take profit of both symmetries exhaustively. The metric can always be 
written, for any of them, as 
\beq 
\label{met-g} 
ds^2 = g_{AB}(R) \, dx^A \, dx^B + G^2(R) \, d\Omega^2, 
\eeq 
where $ A $, $ B = T, R $. $ \partial_T $ has been chosen to be 
the integrable Killing vector and $ d\Omega^2 = d \theta^2 
+ \sin^2 \theta \, d\varphi^2 $. Moreover, if $ G'(R) = 0 $, we already know 
that they belong to the generalized Nariai family, in which case 
they only match with another member of its own family. Thus, we will only 
deal with the situation $ G'(R) \neq 0 $. In this case, a direct 
redefinition of the $ R $ coordinate allows us to write 
\beq 
\label{met-r} 
ds^2 = g_{AB}(r) \, dx^A \, dx^B + r^2 \, d\Omega^2, 
\eeq 
where $ A $, $ B = T, r $. 
 
Spherical symmetry has thus been completely used. We now extract 
consequences from the presence of  $ \partial_T $. The natural thing to do 
is to identify both vector fields, i.e. $ \partial_{T_1} 
\stackrel{\Sigma}{\equiv} \partial_{T_2} $. 
However this is not a right choice, in general, because if a Killing vector 
is multiplied by a constant factor, the resulting vector field is 
obviously a Killing vector field. Therefore, normalizing each 
Killing vector, when possible, gives the natural way to identify 
them. This is indeed implemented in the junction process, if the 
hypersurfaces are {\em spacelike} or {\em timelike} everywhere. On the 
contrary, in the general case, we cannot rely on such normalization. 
 
Any metric of interest (to our purposes) can be written as 
(recall the coordinate change to obtain~\rf{eq-fn0}, 
setting now $ F = 1- H $) 
\beq 
ds^2 = -(1-H) \, dT^2 + {2H \over g} \, dT \, dr 
+ {1+H \over g^2} \, dr^2 + r^2 \, d\Omega^2, 
\eeq 
where $ H $, $ g \neq 0 $ are functions of $ r $ only. 
Looking back to expressions~\rf{met3} and the coordinate changes mentioned 
there,  we will put now $ dT = g_0 dt + ( g_0 - g_0^{-1} ) dr $, where $ g_0 
$ is a constant, that will be related with the function $ g $, as we shall 
see in a moment. With this coordinate change the metric takes the 
form 
\beq 
\label{met-l} 
ds^2 = - g_0^2(1-H) \, dt^2 + 2 {\tilde G} \, dt \, dr 
+ {\tilde F} \, dr^2 + r^2 \, d\Omega^2, 
\eeq 
where $ {\tilde G} = g_0^2 ( H - 1) +1 + H (g_0 - g )/ g $, and 
$ {\tilde F} = 2 + g_0^2 ( H - 1 ) + (g_0 - g)\allowbreak 
[ 2 H / g + 2 / g_0^2 g + ( g_0-g )( H + 1 )/ g_0^2 g ] $. 
 
The junction conditions (for any type of hypersurface, see \cite{tesis9}) 
are then 
\beq 
[ r ] = 0, \quad [{\dot t} ] = 0  \\ 
\label{cond-3} 
\bigl[{\tilde H}\bigr] {\dot t}^2 + 2 \bigl[{\tilde G}\bigr] {\dot t}{\dot 
r} 
+ {\tilde F} {\dot r}^2 = 0, \\ 
\label{cond-4} 
[{\tilde F}] {\dot t}{\ddot r} - [{\tilde G}] ({\ddot t}{\dot t} 
- {\ddot r}{\dot r}) + [{\tilde H}] {\dot r}{\ddot t} 
+ [{\tilde G}' ] {\dot r}^3 + [ {\tilde H}' -({\tilde F}'/2) ] 
{\dot r}^2 {\dot t} - [{\tilde H}' ] {\dot t}^3 / 2 = 0, 
\eeq 
where $ [ f ] \stackrel{\Sigma}{=} f_2 - f_1 $, and where 
we have put $ {\tilde H} \equiv g_0^2( H - 1 ) + 1 $. 
In~\rf{cond-3} and~\rf{cond-4} $ \dot t $ and $ \dot r $ 
are either $ \dot t_1 $, $ \dot r_1 $ or $ \dot t_2 $, $ \dot r_2 $, 
and $ A' \equiv dA(r)/dr|_{r=r(\lambda)} $. 
The same conditions lead, in general, to a second order ordinary 
differential 
equation for $ r $. 
In principle there is the possibility for asymptotic 
stopping solutions, i.e. solutions for which 
$ r \to {\rm const.} $ as $ t \to \infty $, and also for null ones. 
The special case $ r_1 = r_2 = R = {\rm const.} $ is of great interest, 
since it constitutes the solution towards which any transitory solution 
should converge. Under this restriction, the conditions become, simply, 
\beq 
\bigl[{\dot t} \bigr] = 0, \quad 
\bigl[{\tilde H} \bigr] = 0, \quad 
2 \bigl[{\tilde G} \bigr] {\ddot t} - \bigl[{\tilde H}']  {\dot t}^2 = 0, 
\eeq 
where $ t $ is either $ t_1 $ or $ t_2 $. Choosing $ g_0 $ as 
$ g_{\Sigma} $ one gets $ [ {\tilde G} ] = 0 $ 
(the same result comes out directly in the case when the normal 
vector of $ \Sigma $ is non-null). The last conditions 
become then $ [{\tilde H}' ] = 0 $. 
Thus, {\em the conditions emerging 
from the matching of two spherically symmetric spacetimes 
with an integrable Killing vector field are, for the case 
$ r = R = {\rm const.} $ and taking 
the maximum identification between them, 
\beq 
[ \tilde H ] = 0 , \quad [{\tilde H}' ] = 0, 
\eeq 
where $ {\tilde H}  \equiv g_{\Sigma}^2 ( H - 1 ) + 1 $}. An intrinsic 
characterization, valid for any representation of the form~\rf{met-g} 
or~\rf{met-r} (the ones most often dealt with in the literature) 
is $ {\tilde H} \equiv -g^2_{\Sigma} ({\vec \xi} \cdot {\vec \xi}) + 1 $, 
$ g_{\Sigma} \equiv [G' / |\det (g_{AB})|^{1/2}]_{r=R} $, where $ {\vec \xi} 
$ is the Killing vector associated with the staticity of the solution (in 
some regions) of~\rf{met-g} or~\rf{met-r}. Finally, notice that the first 
condition on $ {\tilde H} $ is nothing but the  requirement of the mass 
function to be continuous across the hypersurface, while the second one is 
related with the continuity of the radial stress, or pressure (see 
e.g.~\rf{eq-p}). Needless to say, if one restricts oneself to the family of 
metrics in~\rf{met3}, one gets the conditions of Sect.~\ref{ss-tjo}. 
\section{An application to supersymmetric stringy black \allowbreak holes} 
\label{s-assbh} 
The semiclassical expressions for supersymmetric stringy black holes are 
well-established (see e.g. \cite{horowitz,horpol} and references therein). 
There are also other objects of interest, such as black strings, higher 
dimensional black holes, etc. In all cases, one looks for a correspondence 
principle with general relativistic black holes. This transition is usually 
reflected in the strength of the coupling constant, or the entropy (see e.g. 
\cite{horowitz}--\cite{damour} and references therein). Here we take a 
complementary viewpoint. 
 
The most interesting case to our aims is that of a self-gravitating string 
(see e.g. \cite{damven,damour}). However the necessary ingredients 
---specially the corresponding spacetime metric--- in order to tackle 
this problem are still under study. Here we will consider the most simple 
(and widely considered) case, that of a supersymmetric back hole. 
 
A family of such black holes, related with electrically charged 
black holes, is given by (see \cite{horowitz,horpol} for details) 
\beq 
ds^2 = - f^{-1/2}(r) \Bigl( 1 - {r_0 \over r} \Bigr) 
\, dt^2 + f^{1/2}(r) \biggl[ \Bigl( 1 - {r_0 \over r} \Bigr)^{-1} 
\, dr^2 + r^2 \, d\Omega^2 \biggr], 
\eeq 
where 
$ f(r) = \prod_{i=1}^{4} [ 1 + (r_0 \sinh^2 \alpha_i/r) ] $, and where 
the $ \alpha_i $ are related with the integer charges of the D-branes 
being used. If the correspondence occurs at a constant value of $ r $, we 
get 
\beq 
r_1 + r_0 \sinh^2 \alpha = r_2 f_2^{1/4} (r_2) \equiv R = 
{\rm const.}\\ 
{2m \over R} - {Q^2 \over R^2} = 1 
+ \biggl[\biggl({r_0 \over r} - 1\biggr) 
\biggl( 1 + {r f' \over 4 f} \biggr)^2 \biggr]_{\Sigma_2}, \\ 
- {2m \over R^2} + {2 Q^2 \over R^3} = 
\biggl\{ \biggl( 1 + {r f' \over 4 f} \biggr)^2 
\biggl[{f' \over 2 f}\biggl(1- {r_0 \over r} \biggr) - 
2 {r_0 \over r^2} \biggr] \biggr\}_{\Sigma_2}, 
\eeq 
where we have used $ g_{\Sigma} = G'|_{r = G^{-1}(R)} $, 
$ G(r) = r f^{1/4}(r) $, and $ {\vec \xi} \cdot {\vec \xi} 
= - f^{-1/2} ( 1 - r/r_0) $. The subscript $ \Sigma_2 $ means that all these 
quantities refer to the interior region, to be evaluated at $ r =r_2 $. For 
the exterior metric, we have put $ \alpha_i = \alpha_j \equiv \alpha $, 
for all $ i $, $ j $, because the exterior metric is that of a 
Reissner-Nordstr{\"o}m black hole, for which 
\beq 
\begin{array}{l} 
2m = r_0 \cosh 2 \alpha, \quad Q^2 = r_0^2 \sinh^2 \alpha 
\, \cosh^2 \alpha, \cr 
r_0 = 2 \sqrt{m^2 - Q^2}, \quad 2 \sinh^2 \alpha = -1 
+ m/\!\sqrt{m^2 - Q^2 }, 
\end{array} 
\eeq 
where $ m $ is the (ADM) mass of the black hole and $ Q $ is 
its electric charge. Since $ f_2(r_2) = \prod_{i=1}^4[1 + (r_0 \sinh^2 
\alpha_i/r)]_{\Sigma_2} $, the above conditions yield $ R $ as a function of 
six of the seven parameters, $ M, Q, (r_0)_2, \alpha_i $. 
Detailed analysis shows that these conditions are easily 
fulfilled when $ r_0 \to 0 $, $ \alpha_i \to \pm \infty $, with 
$ r_0 \sinh^2 \alpha_i $ fixed. The resulting $ R $ is very close 
to $ R_0 \equiv m + \sqrt{m^2-Q^2} $, i.e. the event horizon of the 
black hole. We remark that $ r f^{1/4}(r) $ is the radial coordinate 
which has a direct interpretation in terms of the ``size'' of the object, 
and not $ r $ alone. All this being in complete agreement with the 
expected transitions for extreme, and nearly extreme, supersymmetric 
black holes. The same idea should be extended to self-gravitating strings 
when their (4-dimensional) spacetime metric is obtained. For instance, the 
expected order of magnitude of $ R $ found in \cite{damven}, should be 
recovered. This issue will be the  matter of subsequent research. 
\section{Final remarks} 
\label{s-fr} 
The first thing to be noticed is the intrinsic freedom present in our model, 
which is as large as the measure of the set of analytic functions of one 
variable. This is a very rewarding feature, since it allows to impose 
further restrictions coming from new, more accurate proposals. 
In particular, it 
will be a helpful tool when trying to find explicitly a quantum field 
responsible for the $ G_{11} $ and $ G_{22} $ in the fundamental 
uncharged case. For comparison, in all previous works, based on 
individual 
models, the prospective of finding a quantum field related with their 
energy-matter content was hopeless. To that end, we would like to draw the attention to \cite{dB}, where a useful framework to deal with the interior region is given. 
 
In the charged case, let us notice that any GNRSS spacetime can be linked 
with a solution to NED (see \cite{bh02}). Of course, the case of 
Schwarzschild solution is a solution with zero charge and 
Reissner-Nordstr\"om one, the only one which is linear, i.e., Maxwellian. 
Therefore, the {\em whole} family of GNRSS metrics has indeed an immediate 
interpretation in terms of a field theory which is well established when the object is electrically or magnetically charged. This is 
another useful result. A carefull study of this fact will be reported 
elsewhere. 
 
Finally, one can free the requirement that there must be an event horizon. 
The objects would then become ``visible" and the study of the entropy of the 
solutions as well as their associated Hawking temperature would bring some 
clues on the time evolution of (classically) static black holes. 
\section{Conclusions} 
\label{s-c} 
In this work we have investigated, under quite general conditions, the 
question of using Einstein's theory of gravitation ---extended to include 
semiclassical effects--- with the purpose to constraint the physical 
structure of the emerging spacetime solutions that might be suitable for the 
description of the interiors of non-rotating black holes.\footnote{The 
rotating case, which is of major astrophysical interest, and the rotating 
and electrically charged one, which may be associated with spinning 
particles, seem to yield results very similar to the ones presented here, 
see \cite{burinskii},~\cite{behm}.}

In the first part of the work we have exploited the idea that vacuum 
polarization may indeed play an essential role in the interior region. We 
have obtained the result that only two families fulfill the imposed 
requirement and, moreover, we have shown that {\it only one} of them 
is suitable for representing static black hole interiors, what is certainly 
a most remarkable result. 
 
Then we have turned our attention to other sources for the core. Given 
that a promising alternative ---self-gravitating strings---  needs 
still to 
be studied in more detail, we have started this program by first giving 
the general conditions to be fulfilled by any spacetime 
with spherical symmetry and having some static region. Finally, we 
have applied the results obtained  to a supersymmetric black hole, as 
a preliminar case. We have seen that, in such situation, 
the matching is 
generically compatible, including the case of extreme black holes. This last 
setting is precisely the same for which the correspondence between 
semiclassical black holes and stringy ones has been recently confirmed in 
the literature (see e.g. \cite{damven,damour}). 
 
Briefly, our overall conclusion  is the following. First, the results in 
the first sections have opened a new window for the search of a 
compatible quantum field that, once regularized, may yield the same  result 
for, at least, a particular energy-momentum tensor inside the general family 
of models considered (for instance within NED, see also \cite{bh02}). 
Second, once a corresponding 
Einsteinian metric associated with a quantum model is known, the scheme 
developed here has been proven to be well suited to check the 
consistency of 
the involved physical parameters and even, in some cases, to assign explicit 
values to them. 
\section*{Acknowledgements} 
The authors acknowledge valuable discussions with A. Burinskii 
(also his 
careful reading of the manuscript), and with G. Magli. This work has been 
supported by CICYT (Spain), project BFM2000-0810 and by CIRIT (Generalitat 
de Catalunya), contracts 1999SGR-00257 and 2000ACES00017.


\begin{thebibliography}{99} 
\bibitem[*]{email} {elizalde@ieec.fcr.es,  elizalde@math.mit.edu} 
\bibitem[**]{email2}{hildebrandt@ieec.fcr.es} 
\bibitem[$\dagger$]{http} {\tt http://www.ieec.fcr.es/recerca/cme} 
\bibitem{mtw} C.W. Misner, K.S. Thorne and J.A. Wheeler, {\em 
Gravitation} (Freeman, San Francisco, 1973). 
\bibitem{chandra} S. Chandrashekar, {\em The Mathematical Theory of Black 
Holes} (Clarendon Press, Oxford, 1983). 
\bibitem{perl} S. Perlmutter, et al., Nature {\bf 391}, 51 (1998). 
\bibitem{ries}A.G. Riess, et al., Astron. Journ. {\bf 116}, 1009 (1998). 
\bibitem{kottler} F. Kottler, Ann. der Phys. {\bf 56}, 443 (1918); 
Encykl. Math. Wiss. {\bf 22a}, 378 (1922). 
\bibitem{trefftz} E. Trefftz, Math. Ann. {\bf 86} 317 (1922). 
\bibitem{he} S.W. Hawking and G.F.R. Ellis, {\em The Large Scale 
Structure of Space-Time} (Cambridge University Press, Cambridge, 1973) . 
 
\bibitem{casimir} H.B.G. Casimir,  Proc. Kon. Ned. Akad. 
Wetenschap {\bf B 51}, 793 (1948). 
\bibitem{hawking}{S.W. Hawking, Commun. Math. 
Phys. {\bf 55}, 133  (1977).} 
\bibitem{dowker}{J.S. Dowker and R. Critchley, Phys. Rev. 
{\bf D12}, 3224  (1976).} 
\bibitem{hajicek} P. H\'aji\v{c}ek, Phys. Rev.  {\bf D36}, 1065 
(1987). 
\bibitem{mostetru} V.M. Mostepanenko and N.N. Trunov, 
{\em Casimir Effect and Its Applications} (Oxford Science Public., 1997). 
\bibitem{eli2} E. Elizalde, {\em Ten Physical Applications of 
Spectral Zeta Functions}  (Springer, Berlin Heildeberg, 1995). 
\bibitem{eli1} E. Elizalde, S.D. Odintsov, A. Romeo, A.A. Bytsenko 
and S. Zerbini, {\em Zeta regularization techniques with 
applications}  (World Sci., Singapore, 1994). 
 
\bibitem{glinner} E.B. Glinner, Sov. Phys. JETP {\bf 22}, 378 (1966). 
\bibitem{staro} A.A. Starobinsky, Phys. Lett. {\bf 91B}, 99 (1980). 
\bibitem{markov1}M.A. Markov, Ann. Phys. (NY) {\bf 155}, 333 (1984). 
 
\bibitem{ip} E. Poisson and W. Israel, Class. Quantum Grav. {\bf 5}, L201 
(1988). 
\bibitem{fmm} V.P. Frolov, M.A. Markov and V.F. Mukhanov, Phys. Rev.  {\bf 
D41}, 383 (1990). 
 
\bibitem{dymni} I. Dymnikova, Gen. Rel. Grav. {\bf 24}, 235 (1992) 
\bibitem{ds} I. Dymnikova and B. Soltysek, Gen. Rel. Grav. {\bf 30}, 1775 
(1998). 
\bibitem{dymni2} I. Dymnikova, Phys. Lett. {\bf B 472}, 33 (2000). 
\bibitem{ayon} E. Ay\'on--Beato, and A. Garc\'\i a, Phys. Rev. Lett. 
{\bf 80}, 5056 (1998). 
\bibitem{bardeen} J. Bardeen, presented at GR5, Tiflis, U.S.S.R. (1968); A. 
Borde, Phys. Rev.  {\bf D50}, 3392 (1994). 
 
\bibitem{ayon2} E. Ay\'on--Beato, and A. Garc\'\i a, Gen. Rel. Grav., {\bf 
31}, 629 (1999). 
\bibitem{ayon3} E. Ay\'on--Beato, and A. Garc\'\i a, Phys. Lett. 
{\bf B 464}, 25 (1999). 
 
\bibitem{magli1}G. Magli, Rep. Math. Phys. {\bf 44}, 407 (1999) 
 
\bibitem{rqibh} E. Elizalde and S.R. Hildebrandt, ``Regular quantum 
interiors for black holes''. Proceedings of the MG9, Rome (2000); {\tt 
gr-qc/0007030}. 
 
\bibitem{bronnikov}  K.A. Bronnikov, Phys. Rev. {\bf D64}, 064013 (2001). 
 
\bibitem{msi}  R.C. Myers and J.Z. Simon,  Phys. Rev. {\bf D38}, 2434 
(1988). 
\bibitem{wheeler} J.T. Wheeler, Nucl. Phys. {\bf B273}, 732 (1986). 
\bibitem{ks}  R.P. Kerr and A. Schild (1965), in {\it Proceedings  of the 
Galileo Galilei Centenary Meeting on  General Relativity, Problems of Energy 
and Gravitational Waves},  (G. Barbera, ed., Comiato  Nazionale per le 
Manifestazione Celebrative, Florence), pp. 222. 
 
\bibitem{nariai} H. Nariai,  Sci. Rep. Tohoku Univ. Ser. I, 
{\bf 35}, 62 (1951) 
\bibitem{kramer} D. Kramer, H. Stephani, M. MacCallum, and E. Hertl, {\em 
Exact Solutions of Einstein's Field Equations} (Cambridge Univ. Press, 
Cambridge, 1980) p. 156. 
 
\bibitem{mps} M. Mars, M.M. Mart\'\i n-Prats, and J.M.M. Senovilla, 
{Class. Quant. Grav.}, {\bf 13}, L51 (1996); {Phys. Lett.} 
{\bf A 218}, 147 (1996). 
 
\bibitem{israel2} W. Israel, Nuovo Cimento, {\bf 44 B}, 1 (1966); ibid. {\bf 
48 B}, 463. 
\bibitem{robson} E.H. Robson, Ann. Inst. Henri Poincar\'e, Sec. A, XVI(1), 
41 (1972). 
\bibitem{cd} C.J.S. Clarke, and T. Dray, Class. Quantum Grav. {\bf 4}, 265 
(1987). 
\bibitem{marsseno} M. Mars, and J.M.M. Senovilla, Class. Quant. 
Grav., {\bf 10}, 1865 (1993). And references therein. 
\bibitem{sz} W. Shen and S. Zhu, Phys. Lett. {\bf 126A}, 229 (1988). 
 
\bibitem{tesis8} E. Elizalde and S.R. Hildebrandt, {\tt gr-qc/0103063} (from Chap. 8, Ph. D. thesis of S.R.H., University of Barcelona, 2001). 
 
\bibitem{astro} M.J. Rees, in {\em BHs and Relativistic Stars}, ed. 
R.M. Wald (The University of Chicago Press, Chicago, 1998). 
 
\bibitem{tolman} R.C. Tolman, {\em Relativity, thermodynamics and Cosmology} 
(Clarendon Press, Oxford). 
 
\bibitem{bp} R. Balbinot and E. Poisson, Phys. Rev. {\bf D41}, 395 (1990). 
 
\bibitem{burinskii} A. Burinskii,``{Supersymmetric superconducting 
bag as a core of Kerr spinning particle}'', {\tt hep-th/0008129}. 
Also in the Proceedings of the MG9, Rome (2000). 
\bibitem{morris} J.M. Morris, Phys. Rev. {\bf D 56}, 2378 (1997). 
\bibitem{cvs} M. Cveti\v{c} and H. Soleng, Phys. Rep. {\bf 282}, 159 
(1997); M. Cveti\v{c}, S. Griffies, and  S.J. Rey, Nucl. Phys. 
{\bf B381}, 301 (1992). 
 
\bibitem{borde} A. Borde, Phys. Rev.  {\bf D55}, 7615 (1997). 
 
\bibitem{bm} J. Bekenstein and A. Mayo, Phys. Rev. {\bf D54}, 5059 (1996). 
 
\bibitem{visser} M. Visser, ``{Gravitational vacuum polarization}'', 
in {\em Proceedings 8th Marcel Grossman Meeting}, (Jerusalem, 1998). 
 
\bibitem{bd} N.D. Birrel and P.C.W. Davies, {\em Quantum Theory in Curved 
Spaces} (Cambridge University Press, Cambridge, England, 1982); A.A. Grib, 
S.G. Mamayev and V.M. Mostepanenko, {\em Vacuum Quantum Effects in Strong 
Fields} (Friedman Laboratory Publishing, St.Petersburg, Russia, 1994). 
 
\bibitem{tesis9} E. Elizalde and S.R. Hildebrandt, {\tt gr-qc/0103064} 
(from Chap. 9, Ph. D. thesis of S.R.H., University of 
Barcelona, 2001). 
 
\bibitem{horowitz}  G.T. Horowitz, in  {\em Black Holes and Relativistic 
Stars}, ed. R.M. Wald, (The University of Chicago Press, Chicago, 1998). 
\bibitem{horpol} G.T. Horowitz and J. Polchinski, Phys. Rev. {\bf D57}, 
2557 (1998). 
 
\bibitem{damven} T. Damour and G. Veneziano, Nucl. Phys. {\bf B568}, 
93 (2000). 
\bibitem{damour}  T. Damour, ``Quantum strings and black holes''. 
Proceedings of the MG9, Rome (2000); {\tt gr-qc/0104080}. 
 
\bibitem{dB} A. DeBenedictis, D. Aruliah, and A. Das, Gen. Rel. 
Grav., to be published; gr-qc/0105123. 
 
\bibitem{behm} A. Burinskii, E. Elizalde, S.R. Hildebrandt, and 
G. Magli, 
Phys. Rev. {\bf D65}, 064039 (2002); gr-qc/0109085. 
 
\bibitem{bh02}  A. Burinskii and S.R. Hildebrandt, Phys. Rev. D, to be 
published; hep-th/0202066. 
 
\end{thebibliography}
\end{document}